\documentclass[12pt]{article}

\usepackage{scicite}
\usepackage{graphicx}

\usepackage{times,amsmath,amssymb}

\newenvironment{sciabstract}{%
\begin{quote} \bf}
{\end{quote}}

\newcounter{lastnote}
\newenvironment{scilastnote}{%
\setcounter{lastnote}{\value{enumiv}}%
\addtocounter{lastnote}{+1}%
\begin{list}%
{\arabic{lastnote}.}
{\setlength{\leftmargin}{.22in}}
{\setlength{\labelsep}{.5em}}}
{\end{list}}

\title{The Triple-Ring Nebula around SN 1987A: Fingerprint of
a Binary Merger}

\author
{Thomas Morris,$^{1,2}$ Philipp Podsiadlowski$^{1,\ast}$\\ 
\\
\normalsize{$^{1}$Dept.\ of Astrophysics, University of Oxford, Oxford
OX1 3RH, UK}\\
\normalsize{$^{2}$Max-Planck Institut f\"ur Astrophysik, Garching, Germany}\\
\normalsize{$^\ast$E-mail:  podsi@astro.ox.ac.uk.}
}

\date{}

\begin{document} 

% Double-space the manuscript.

%\baselineskip24pt

% Make the title.

\maketitle

% Place your abstract within the special {sciabstract} environment.

\begin{sciabstract}
Supernova 1987A, which was the first naked-eye supernova since
Kepler's supernova in 1604. It was an unusual supernova that in many
respects defied theoretical expectations. It has long been suspected
that these anomalies were the result of the merger of two massive
stars some 20,000 years before the explosion, but so far there has
been no conclusive proof. Here we present three-dimensional
hydrodynamical simulations of the mass ejection associated with such a
merger and the subsequent evolution of the ejecta and show that this
reproduces the properties of the triple-ring nebula surrounding the
supernova in all its details.
\end{sciabstract}

Supernova 1987A (SN 87A) in the Large Magellanic Cloud was one of the
major astronomical events of the 1980s, but it was highly unusual.
The progenitor star, Sk $-$69$^{\circ}$202, was one of the
major surprises. While massive stars similar to the progenitor
of SN~87A are expected to end their evolution as red supergiants,
Sk $-$69$^{\circ}$202 was a blue supergiant. Moreover,
the outer layers of the star were highly enriched in
helium {\it (1)}, suggesting that some nuclear processed
material from the core was mixed into the envelope by a non-standard
mixing process {\it (2)}. Most spectacularly, the supernova is
surrounded by a complex triple-ring nebula {\it (3, 4)}, 
consisting of material that was ejected from the progenitor some
20,000\,yr before the explosion in an almost axi-symmetric but in a
very non-spherical manner. All of this has pointed to some dramatic
event that occurred to the progenitor some 20,000\,yr before the
explosion, most likely the merger of two massive stars {\it (5)}.

A merger was first suggested to explain some of the asymmetries of
the supernova ejecta {\it (6)}. Later it was realized that a
binary merger can also explain the blue progenitor and the main
chemical anomalies of the event {\it (7\,--\,9)}. The 
latter was confirmed in detailed stellar, hydrodynamical simulations
of the slow merger of two massive stars {\it (10)}. However, the
origin of the triple-ring nebula has so far not found a satisfactory
explanation.

The triple-ring nebula consists of three overlapping rings in
projection.  The supernova occurred at the center of the inner ring,
while the outer rings are in planes almost parallel to the central
ring plane but displaced by 0.4~pc above and below the central ring
plane.  It is important to note that these outer rings do not form the
limb-brightened projection of an hourglass nebula, as in some of the
early models for the nebula {\it (11\,--\,13)}, but dense,
ring-like density enhancements {\it (4)}. Previous attempts to
model the nebula have involved interacting winds in a
binary {\it (14, 15)}, a photoionization-driven
instability {\it (16)}, mass ejection during a binary
merger {\it (17)} or magnetically controlled ejection {\it (18)},
but none of these have been able to fully explain both the detailed
geometry and the kinematical properties of the nebula. 

The axi-symmetric, but very non-spherical structure immediately
suggests that rotation may have played a role in the formation.
However, it follows from simple angular-momentum conservation
arguments that any single star that is rotating rapidly in its early
evolution could only be rotating slowly after it has expanded by a
factor of $\sim 100$ to red-supergiant dimensions.  On the other hand,
a binary merger provides a simple and effective means of converting
orbital angular momentum into spin angular momentum and of producing a
single, rapidly rotating supergiant.

In a typical binary merger model for SN~87A {\it (9)}, the system
initially consisted of two massive stars with masses of $\sim
15\,M_\odot$ and $\sim 5\,M_\odot$ in a fairly wide orbit with an
orbital period longer than $\sim 10\,$yr, so that the more massive
component only starts to transfer mass to its companion after it has
completed helium burning in the core. Because of the large mass ratio,
mass transfer is expected to be unstable and to lead to a common
envelope phase, where the secondary is engulfed within the primary's
envelope (Fig.~1). In this phase, most of the orbital angular momentum
of the binary is deposited in the common envelope spinning it up in
the process. Because of friction with the envelope, the secondary will
start to spiral-in inside the common envelope. Since the density
increases inwards, the spiral-in process initially accelerates, until
enough orbital energy has been released and deposited in the envelope
to affect the envelope dynamically.  Because the timescale of this
energy ejection is comparable to the dynamical timescale of the
envelope, the energy injection into the envelope is essentially
impulsive leading to a dynamical expansion of the envelope and
significant mass ejection. After the envelope has expanded, the
spiral-in of the secondary slows down and now occurs in a
self-regulated manner, where all the orbital energy released is
transported to the surface, where it is radiated away. Ultimately, the
secondary inside the envelope will merge with the core leading to the
dredge-up of core material from the primary core {\it (10)}. The
whole merger process is expected to take a few 100\,yr. Afterwards the
star will initially be an oversized red supergiant, but will shrink in
a few 1000\,yr, the timescale on which the envelope loses its excess
thermal energy, to become a blue supergiant. In this latter phase, the
fast, energetic wind from the blue supergiant will sweep up the ejecta
associated with the merger and shape the whole nebula.

To simulate the mass ejection during the merger and the subsequent
evolution of the ejecta, we use GADGET {\it (19)},
a three-dimensional, particle-based hydrodynamics code,
using the smooth-particle hydrodynamics (SPH) method. We split the
simulations into two parts. First, we simulate the mass ejection
associated with the merger and then take the output from this model
to start a second calculation to simulate the subsequent evolution
of the ejecta as they are being swept up by the blue-supergiant wind.

We model the red supergiant as a polytrope with a central point source
of $8\,M_\odot$, representing the compact core of the star and the
immersed companion, and an envelope mass of $12\,M_\odot$; the initial
radius of the star is taken to be $1500\,R_\odot$.  We then mimic the
initial spin-up phase by adding angular momentum to the envelope over
a period of $\sim 6\,$yr until all the angular momentum from the
initial binary is deposited in the envelope {\it (20)}. Because of
this spin-up, the envelope will become highly non-spherical and take
on a disk-like shape (Fig.~2a). Most of the orbital energy is injected
impulsively when the orbital energy released is comparable to the
binding energy of the envelope {\it (21)}. This produces an inner
region of overpressure which starts to expand and drive a shock in the
overlying layers, ejecting some of the outer layers in the
process. Because of the highly flattened envelope structure, mass
ejection is easiest in the polar direction and occurs there
first. Whether mass is ejected in the equatorial direction depends on
the amount of energy injected. If it is less than $\sim 1/3$ the
binding energy of the envelope, the large mass concentration in the
equatorial plane inhibits equatorial mass ejection {\it (20)}:
particles that try to escape in the equatorial direction are deflected
by the large equatorial mass concentration towards higher latitudes
(Fig.~2c). This produces a large density enhancement in the ejecta at
roughly 45 degrees. Indeed, it is the overdensity at mid-latitude that
will ultimately be swept up to form the outer rings in the SN~87A
nebula.

In our best model, no matter is ejected in the equatorial
direction. However, because the merged object has much more angular
momentum than a more compact blue supergiant could have, the merged
object has to lose this excess angular momentum, most likely in the
form of a slow equatorial outflow {\it (22)}, as it shrinks to become
a blue supergiant. We estimate that, for typical parameters, most
likely several solar masses need to be lost in this transition phase
(see {\it (21)} for further discussion). Once the merged object has
become a blue supergiant, its energetic blue-supergiant wind will
start to sweep up all the structures ejected previously.

To model the blue-supergiant phase, we start a second SPH calculation
where we only simulate the ejecta. For the initial model of this
calculation we take the output from the first simulation once the
ejecta expand freely. This gives the mass and the velocity of the
ejecta as a function of latitude. We then ballistically extrapolate
the evolution of the ejecta for $4000\,$yr, the assumed time of the
red--blue transition. We model the equatorial mass shedding by
including an equatorial outflow, lasting for $2000\,$yr {\it (21)}.
We then turn on a spherically symmetric blue-supergiant
wind (with a mass loss rate of $\dot{M}= 2\times
10^{-7}\,M_\odot\,$yr$^{-1}$ and wind velocity $v_{\rm w} =
500\,$km\,s$^{-1}$) and follow the subsequent evolution.

In a typical simulation after $\sim 20,000\,$yr (Fig.~2d), the density
enhancement at mid-latitude has been swept up into two well-defined
rings, which together with the swept-up equatorial outflow produce the
main features of the triple-ring nebula when observed at an
inclination of $\sim 45$\,deg (Fig.~2e).

The HST image {\it (4)} of the nebula shows that the symmetry
center of the outer rings is slightly displaced from the symmetry axis
of the central ring. This asymmetry can be easily explained if the
ejecta associated with the merger are given a small kick of $\sim
2\,$km\,s$^{-1}$ in a direction to the north-west of the nebula, at an
angle of $11^{\circ}$ out of the equatorial ring plane.  This was done
to produce the model shown in Figure~2e, which almost perfectly
reproduces not only the main features of the triple-ring nebula but
also the small asymmetries of the outer rings (their deviations from
perfect ellipses and their displacement from the central symmetry
axis).  To generate this emission model {\it (21)}, we assumed that
the nebula is fully ionized by the SN explosion and that the emission
is optically thin. This `best' model also reproduces the kinematics of
both the inner ring {\it (23)} and the outer
rings {\it (4)}.  In this particular model, the inner ring contains
$0.4\,M_{\odot}$ of mass, while the outer rings contain a total of
$0.02\,M_{\odot}$ each.

The physical origin of this small kick is not entirely clear. It could
be associated with a non-radial pulsational mode excited during the
early spiral-in phase; alternatively it could be due to the orbital
motion caused by a more distant low-mass third star in the system.

In addition to the three rings, the model predicts several other
structures. The outer rings form the ends of two bipolar lobes which
combined with the bow-shock structure around the inner ring are
reminiscent of an hourglass nebula (Fig.~2d). This structure is
presumably the origin of the hourglass structure that has been
inferred from light-echo studies {\it (24)}. Further light echoes
were also detected from beyond the triple-ring structure as summarised
in {\it (25)}; these were most likely formed during the
earlier red-supergiant phase of the primary star before the merger or
in the phase immediately preceding the merger, where accretion on
the companion may produce a bipolar outflow (e.g. {\it (26)}), 
phases not included in our simulations.

The hydrodynamical model we have presented here provides an excellent
fit to the observed triple-ring nebula around SN~1987A.  It is
important to emphasize that this model does not require any physically
ad hoc assumptions and that all the input parameters are
compatible with the values expected from simple modelling of the
various phases of the merger (apart from the physical origin of the
small kick to the ejecta, which is presently not explained).

The model also makes several predictions, specifically about the mass
in the different rings and other structures. These may ultimately
become visible when the nebula is being destroyed by the supernova
ejecta or when the nebula is being re-ionized by the emergent X-ray
flux resulting from the supernova--ring interaction.  Since in our
favoured model the outer rings are ejected before the dredge-up of
core material, while the inner ring is ejected afterwards, one may
expect some chemical differences between the inner ring and the outer
rings, where the inner ring should show a larger helium enhancement
and more evidence for CNO processing than the outer rings. There is
the tantalizing hint that this may already have been
observed {\it (27)}, but this needs to be confirmed in a more
detailed comparative study.

\goodbreak

{\obeylines\parindent=30pt
{\bf Supporting Online Material}
www.sciencemag.org
Materials and Methods
Nebula Geometry
Tables S1, S2
Movie S1, S2
}

\bigskip\bigskip\bigskip

\def\ga{\mathrel{{\lower3pt\hbox{$\sim$}}\hspace{-10pt}
     \raise2.0pt\hbox{$>$}}}

\parbox{15cm}{{\bf Figure 1.} Schematic diagram illustrating the
formation of the triple-ring nebula. The system initially consisted of
a binary with two stars of $\sim 15\,M_\odot$ and $\sim
5\,M_\odot$, respectively, with an orbital period $\ga 10\,$yr.  (a) Mass
transfer is dynamically unstable leading to the merger of the
two components in a common envelope and (b) the spin-up of the envelope.
(c) The release of orbital energy due to the spiral-in of the
companion leads to the partial ejection of the envelope.
(d) After the merging has been completed, the merged object evolves to
become a blue supergiant, shedding its excess angular momentum in an
equatorial outflow.  In the final blue-supergiant phase, the energetic
wind from the blue supergiant sweeps up all the previous structures,
producing the triple-ring nebula.}
\vfill\eject

\phantom{nothing}
\begin{figure}[h]
\centering
\includegraphics[width=15cm,angle=0]{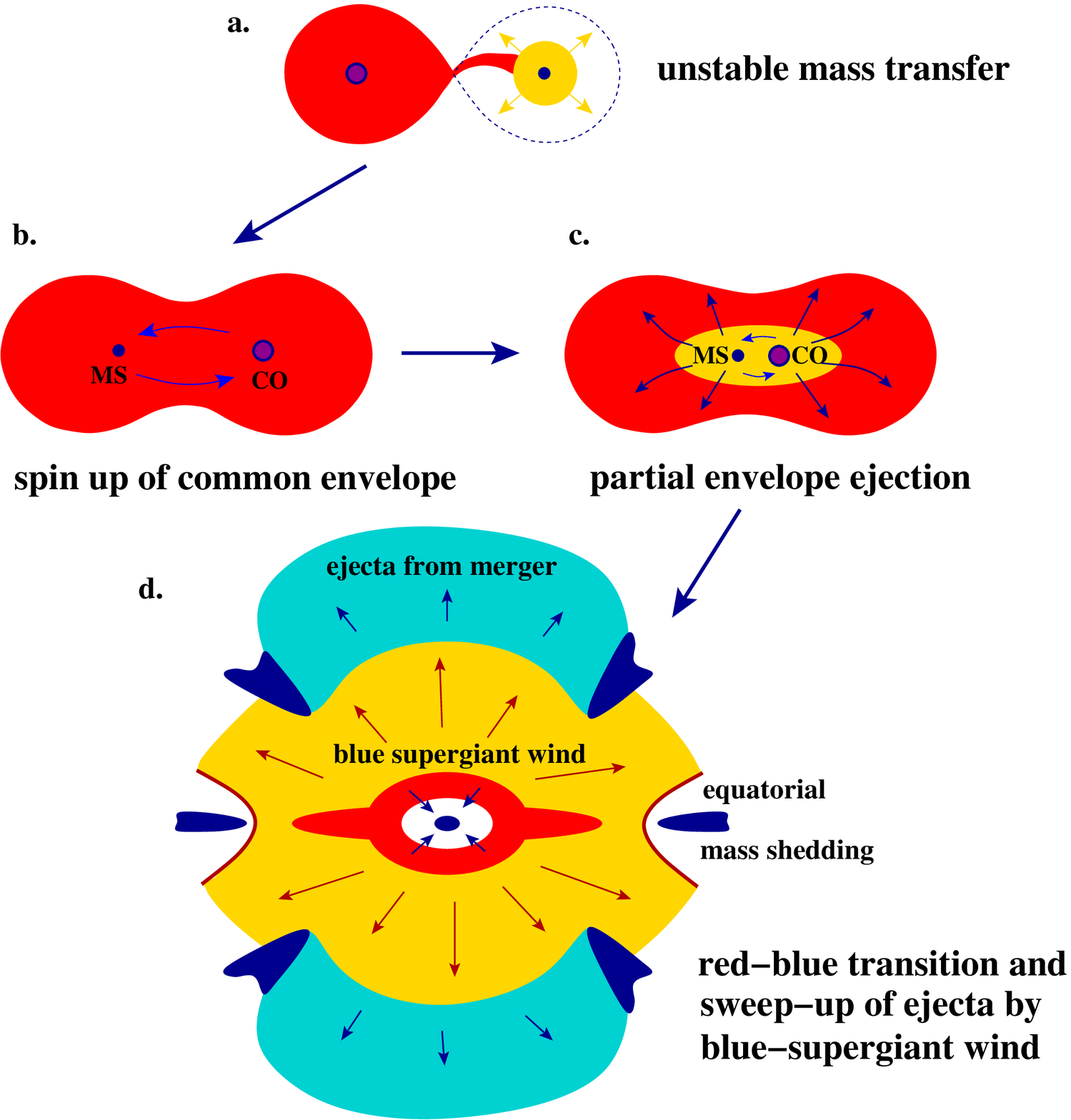}
\end{figure}

\vfill\eject

\parbox{15cm}{{\bf Figure 2.}
Three-dimensional hydrodynamical (SPH) simulations to model the various phases
illustrated in Fig.~1.
{(a):} Cross-section of the rapidly rotating red-supergiant
envelope, containing $2\times 10^5$ SPH particles, following the
spin-up in the initial common-envelope phase (Fig.~1b) after a total
amount of orbital angular momentum of $L=8\times
10^{54}\,\mathrm{erg\,s}$ has been deposited in the envelope. In these
units, the red supergiant had an initial radius of 1.
{(b):}
Particle snapshot in the meridional plane approximately $3$\,yr after the
orbital energy has been injected in the central part of the common
envelope (cf.\ Fig.~1c), showing the formation of the enhancement at
mid-latitudes. Red particles are unbound.
{(c):} The mass enhancement in the ejecta plotted as a function of
time (contours) and distance from the centre of mass, as a function of
latitude. The horizontal axis shows the equatorial plane where no mass
is ejected, and the colour scale beneath the plot shows the increase
of ejected mass over the value expected in spherical symmetry. The
contours show the mean radius of the ejected material at the time shown,
in units of $0.8\,\mathrm{yr}$ {\it (21)}.
The axes are in the same unit system as those of (a) and (b).
{(d):} The final particle distribution ($10^6$ particles)
plotted in the meridional plane at an age of 20\,kyr. The density and
mean velocity of the material in the equatorial ring (outer rings) is $\sim
2\times 10^4\,\mathrm{cm^{-3}}$ ($10^3 \mathrm{cm^{-3}}$) and 10.3
$\mathrm{km\,s^{-1}}$ (31 $\mathrm{km\,s^{-1}}$),
respectively. Wind particles are shown in black while the nebula
particles are shown in dark blue through pale green based on a
logarithmic density scale as indicated (in CGS units).  The axes are
in units of $3\times 10^{18}\,\mathrm{cm}$.
{(e):} Simulated emission measure at $\sim 2000\,$d after the supernova
in the 656nm H$\alpha$ line for our `best' model. The
total flux from the outer rings is $4\times 10^{45}\,\mathrm{photons\,s^{-1}}$,
comparable to the observed flux {\it(4)}.}
\vfill\eject

\begin{figure}
 \centering
  \includegraphics[height=20cm,angle=0]{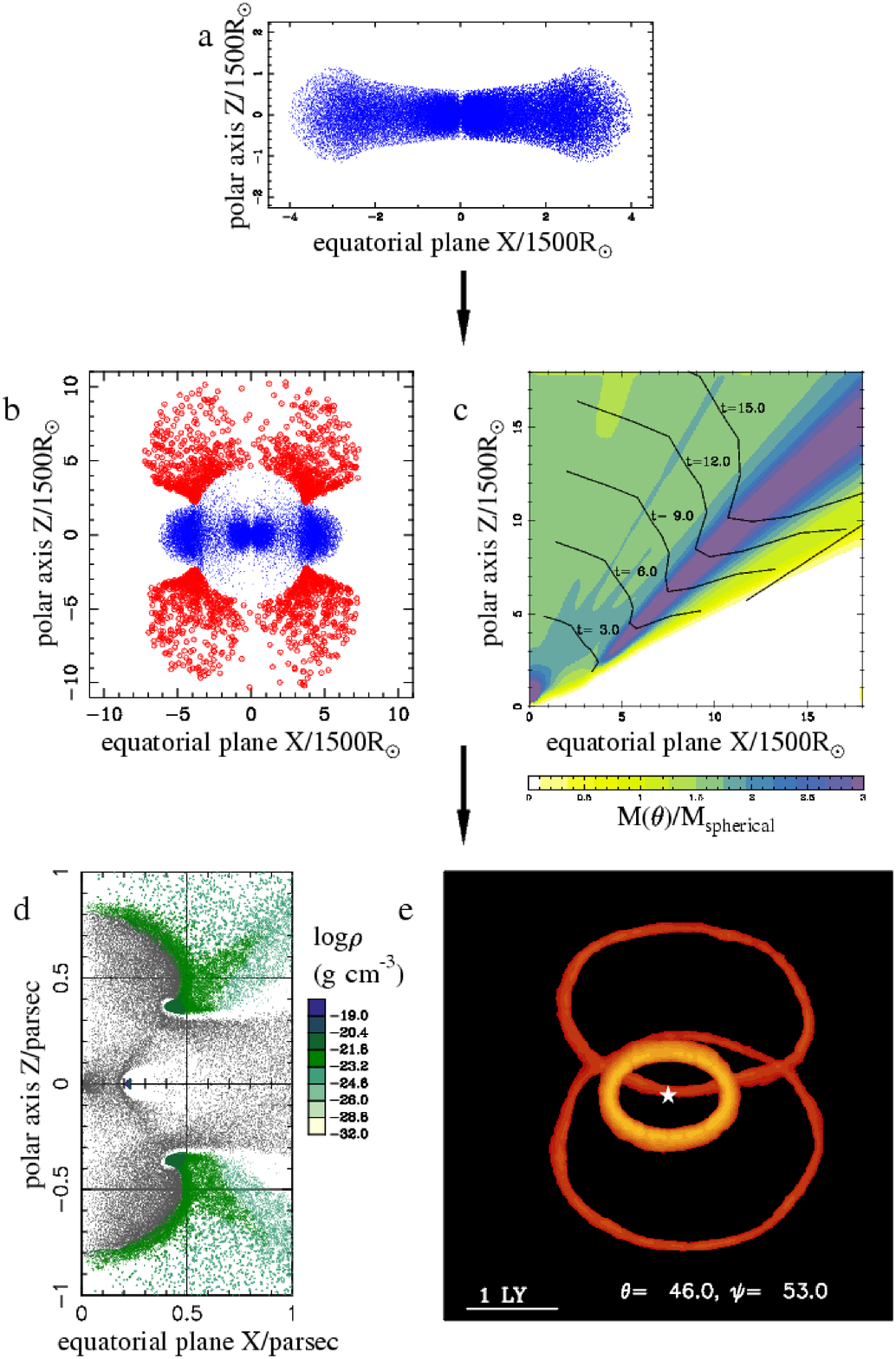}
\end{figure}

\clearpage
\section*{Materials and Methods}
\subsection*{Binary merger phase}

All hydrodynamical simulations were performed with a modified version
of the publicly available GADGET code {\it(S1)}, which uses the 
smooth-particle-hydrodynamics (SPH) method {\it(S2)} to simulate
hydrodynamical flows of arbitrary geometry in three dimensions.

At the onset of the common envelope phase, the primary star is an
evolved red supergiant which we approximate as a polytrope
with polytropic index $n=3/2$ containing a centrally condensed core of
$M\approx 2/5M_{\star}$. The spherically symmetric (non-rotating)
density profile is sampled with $2\times 10^5$ particles using a
Monte-Carlo technique which are then relaxed using the method of Lucy
{\it(S3)} to reduce particle noise. An adiabatic equation of state
$\gamma=5/3$ is used throughout, except for the relaxation phase for
which constant entropy $s\propto p/\rho^{\gamma}$ is assumed ($p$ and
$\rho$ are the pressure and density, respectively).
The envelope contains significant mass at
large radii and is therefore able to store a large fraction of the
available orbital angular momentum
\begin{equation}
  L_{\rm orb}=
  6.60 \times 10^{54} \mathrm{g\,cm}^2 \mathrm{s}^{-1} \,
       A_{2500}^{1/2} \,
       M_{15} \, M_5 \, M_{20}^{-1/2},
  \label{eq:angmom}
\end{equation}
which is transferred to the envelope during the early
spiral-in of the companion.
Here $A_{2500}$ is the orbital separation in units of
2500\,$R_{\odot}$, $M_{15}$ and $M_{5}$ are the masses of the primary
and the secondary in units of $15\,M_{\odot}$ and $5\,M_{\odot}$ (as
indicated by the subscripts), respectively, and $M_{20}=M_1+M_2$ is
the total mass in units of $20\,M_{\odot}$.
Previous one-dimensional calculations including energy and
angular momentum transport {\it(S4, S5)}
suggest that most angular momentum is deposited in the early spiral-in
phase.

In our model, the angular momentum is added to the envelope slightly
more slowly than the sound-crossing time of the envelope
in such a manner that the particles remain
sub-Keplerian at all times. The gain in angular velocity on each
particle is given by
\begin{equation}
  \Delta\Omega =
  \begin{cases} \gamma \Delta t & \text{if $v_i/v_{\mathrm{Kepler}}<1$
      always,} \\
    0 & \text{otherwise}
  \end{cases}
\end{equation}
which leads to a stable angular momentum profile with a core rotating
close to the local Keplerian velocity at each cylindrical radius, 
and an extended envelope with a
constant specific angular momentum profile (see Fig.~1 of {\it(S6)}).

The energy deposited in the envelope corresponds to the orbital energy
lost by the companion during the plunge-in phase when its orbit
decreases from $r\sim 1R_{\star}$ to $r\sim 0.1R_{\star}$. The
available energy is roughly (also see table~S1)
\begin{equation}
  \Delta E_{\rm orb}= 5.0\times 10^{46}\,\mathrm{erg}\, M_{10}M_2\,(R_i/R_f-1),
\end{equation}
where $M_{10}$ and $M_2$ are the primary and secondary masses at a
separation of $\sim 1/5R_{\star}$ in units
of 10 and 2 solar masses, respectively, and $R_i$ and $R_f$ are the
initial and final orbital separation. Most of this energy is deposited
in the inner envelope, corresponding to $1.5\times 10^{47}$\,erg in
the merger model discussed in this paper, or half of the available orbital
energy. The deposited energy is added to the
central region $r<200\,R_{\odot}$ which contains $0.7\,M_{\odot}$,
corresponding to a temperature of $9\times 10^5$\,K. We have tested
that the exact details of the energy deposition does not strongly affect
the amount of mass in the ejecta and their geometry (also see {\it(S6)}).

The latitude-dependence of the outflowing material is illustrated in
Fig.~2c in the paper. The image shows the mass enhancement ratio
$M(r,\theta)/M_{\mathrm{spherical}}$ where $M(r,\theta)$ is the
ejected mass in a spherical coordinate system, divided by the
spherically symmetric value $M_{\mathrm{spherical}}$. The horizontal axis
lies in the equatorial plane while the vertical axis lies along the
polar direction. The contours show the median radius $r(t)$ of the ejected
material at a given time $t$ (labelled) in units of 0.8\,yr, as a
function of latitude, showing that the highest velocity material is
ejected in the polar direction. The mid-latitude enhancement (coloured
purple) corresponds to a local velocity minimum, and no material is
ejected in the equatorial plane (white,
$M(r,\theta)/M_{\mathrm{spherical}}<0.2$).

\subsection*{The blue loop}
Following the merger, the star contracts by a factor $\gtrsim 10$ in
radius and becomes a blue supergiant, on a timescale of a few
thousand years. The stellar wind of the star changes from a slow dense
outflow ($v_w\sim 30\,\mathrm{km\,s^{-1}}$) to a much faster wind
($v_w\sim 500-800\,\mathrm{km\,s^{-1}}$) which compresses the dense gas
ejected during the binary merger. Typically 10 per-cent
of the envelope is ejected, containing $2\times 10^4$ particles which
are resampled to increase the resolution to $\sim 10^6$ particles,
preserving the density enhancement at mid-latitudes. These new particles
are integrated on ballistic trajectories for 4\,kyr to set the initial
gas distribution at the beginning of the blue-supergiant phase.

The equatorial outflow occurs as the envelope loses angular
momentum during the blue loop. The total mass lost can be estimated
from angular momentum conservation
\begin{equation}
  M_{ER}= \frac{\Delta L}{\sqrt{GM_{\star} R_{\star} (1-\Gamma)}}
   \sim 4\,M_{\odot}
\end{equation}
if the mass is lost near critical rotation (as suggested by the low
expansion velocity of the ring of around $10.3\,\mathrm{km\,s^{-1}}$).
Here $\Delta L$ is the excess angular momentum that needs to be lost, i.e.
is the difference between the post-merger angular momentum in the
envelope and the maximum angular momentum for a stable blue supergiant
($\sim 4\times 10^{54}{\,\rm erg\,s}$).  We assume an Eddington factor
of $\Gamma=0.4$ and that the envelope must lose $\sim 6\times
10^{54}$\,erg\,s at a radius of roughly $6000\,R_{\odot}$. This is
likely to be a lower limit unless magnetic processes in the excretion
disk can increase the specific angular momentum of material at larger
radii. 

We model the equatorial mass shedding by including an outflow, lasting
for $2000\,$yr, with an assumed angular density profile $\rho \propto
\cos^2 \left({\pi\theta}/{2\,\theta_0} \right)$ ($|\theta| \le
\theta_0$, where $\theta$ is the angle relative to the equatorial
plane and $\theta_0$ is typically 5 degrees) and the ring contains of
order $0.5\,M_{\odot}$.

We note that, while the existence of the inner ring is important 
for the shaping of the outer rings (see the bow shock near the inner
ring in Fig.~2d), the shape of the outer rings is not sensitive to
the amount of mass or the detailed geometry of the equatorial
outflow.

\subsection*{Nebular phase}
During the final blue supergiant phase, the supersonic stellar wind
sweeps up the pre-existing material. In order to model this phase, we
have modified the GADGET code to include the addition of new gas
particles at small radii, relative to the shell, to represent the fast
wind. The acceleration of these wind particles can be neglected since
their initial radius $r_0$ is much larger than the typical stellar
radius: $r_0\sim 10^{-3}\,\mathrm{pc} \sim 10^3R_{\star}$.

At each interval $\delta t$ which is small relative to the
crossing time of the free wind region, new particles are
distributed in a sphere with radius $v_w\delta t$ where
$v_w=500\,\mathrm{km\,s^{-1}}$ is the terminal velocity of the
blue supergiant wind. The radial profile follows $r^{-2}$ in the free
wind region. A fixed number of particles (typically $N_w \sim 500$) are
introduced, with equal masses calculated from the stellar mass-loss
rate, which is assumed constant throughout the duration of the blue
supergiant phase:
\begin{equation}
  N_w m_w = \dot{M} \delta t
\end{equation}
These are treated as gas particles in the smoothed particle
hydrodynamics-based code. The code has been verified against the
semi-analytic solution in the case of an adiabatic bubble in a
constant density medium {\it(S7)}.

\subsection*{Flash ionization by the supernova explosion}
In order to compare our results to the {\small HST} image
we simulate the H$\alpha$ emission from the nebula by integrating the
square of the density along the line of sight. The relative pixel
intensity is simply $\int n_e n_{H^+} \mathrm{d}l= \int n_H^2
\mathrm{d}l$ if the entire nebula is initially
ionized. A simple recombination model is included with the form
$n_{H^+}(t)=n_H(0) \exp(-\alpha n_H t)$ where $n_H$ and $n_{H^+}$ are the
total hydrogen
density and proton density, respectively, and $\alpha_B\approx 3\times
10^{-13}\mathrm{cm^3\,s^{-1}}$ is the hydrogen
recombination coefficient to levels $n\ge 2$. For
simplicity we assume $n_e=n_{H^+}$ at all times. The pixel intensity
due to a narrow ring or filament of dense material is thus
\begin{equation}
  I\propto n_{H^+}^2 \Delta r,
\end{equation}
which gives $I\sim 5\times 10^{24}\,\mathrm{cm^{-5}}$ for the
equatorial ring and $I\sim 2-5\times 10^{22}\,\mathrm{cm^{-5}}$ for
the outer rings.  Pixels below a threshold intensity of $I= 2\times
10^{21}\,\mathrm{cm^{-5}}$ along the line of sight are set to zero
intensity. The equatorial ring is probably only ionized to a
skin-depth of around $\sim 4-10\times 10^{15}\,\mathrm{cm}$
{\it(S8)} while the rest of the ring remains neutral. The
hydrogen (and much of the helium) in the outer rings is expected to be
completely ionized by the supernova outburst.  The density in the
outer rings is found to be 100 to 150 times the density in the shocked
blue-supergiant wind bubble and is a factor of 5 to 10 larger than the
density in the polar lobes, accounting for the sharp appearance of the
outer rings.

\section*{Supporting Text}
\subsection*{Nebula geometry}
The latitude of the outer rings is determined primarily by the
location of the mass peak in the merger ejecta. The
ratio of the radius of the equatorial ring to that of the outer loops
is determined firstly by their
velocities of ejection, which are unconstrained in the model, and
secondly by their relative masses. Values of $2.5-3.5$ are typical for
model parameters applicable to SN\,1987A (table~S2).

\newpage
\section*{Supporting Tables}

\begin{table}[h]
 \centering
  \caption{Selected properties of the polytope model before the
  spin-up/spiral-in phase (left column) and after the addition of $L=8
  \times 10^{54}$ erg\,s of angular momentum (right column).}
  \vspace{15pt}
  \begin{tabular}{lcc}
  \hline
  \noalign{\vspace{2pt}}
  Property & Before spin-up & After spin-up \\
  \noalign{\vspace{2pt}}
  \hline
  \noalign{\vspace{2pt}}
  Potential energy $W/10^{47}$ erg & $-12.0$ & $-7.2$ \\
  Kinetic energy $T/10^{47}$ erg   & 0     & 1.1  \\
  Binding energy/ $10^{47}$ erg    & $-6.2$  & $-4.5$ \\
  Ratio $T/W$                      & 0     & 0.15 \\
  Angular momentum $L/10^{54}$ erg\,s             & 0   & 8.0  \\
  Moment of inertia $I/10^{61} \mathrm{g\,cm^2}$  & 4.1 & 73.6 \\
  \noalign{\vspace{2pt}}
  \hline
  \end{tabular}
\end{table}

\bigskip\bigskip
% Ratios of the radius of the inner/outer rings
\begin{table}[h]
 \centering
  \caption{Values of the radius ratio of the outer rings to the
  equatorial ring, for representative SN\,1987A
  nebula calculations at an age of 22\,kyr after the onset of the 
  blue-supergiant wind. The units of the initial equatorial ring velocity
  $v_{ER,0}$, mass $M_{ER}$ and wind mass-loss rate $\dot{M}$ are
  $\mathrm{km\,s^{-1}}$, $M_{\odot}$ and
  $10^{-7}\,M_{\odot}\,\mathrm{yr^{-1}}$, respectively.}
  \vspace{15pt}
  \begin{tabular}{lccccc}
  \hline
  \noalign{\vspace{2pt}}
  Model & $v_{ER,0}$ & $M_{ER}$ & $\dot{M}$ & $r_{NOR}/r_{ER}$ &
  $r_{SOR}/r_{ER}$ \\
  \noalign{\vspace{2pt}}
  \hline
  \noalign{\vspace{2pt}}
  M1    & 8.0      & 0.4      & 2.0       & 2.79 & 2.86 \\  % 2.11
  M2    & 8.5      & 0.4      & 2.0       & 2.58 & 2.65 \\  % 2.14
  M3    & 9.0      & 0.4      & 2.0       & 2.38 & 2.44 \\  % 2.17
  M4    & 8.0      & 0.6      & 2.0       & 3.03 & 3.08 \\  % 2.29
  M5    & 8.0      & 0.8      & 2.0       & 3.09 & 3.13 \\  % 5.22
  M6    & 8.0      & 1.0      & 2.0       & 3.15 & 3.17 \\  % 5.37
  M7    & 8.0      & 1.0      & 2.5       & 3.27 & 3.32 \\  % 5.38
  M8    & 8.0      & 1.0      & 3.0       & 3.38 & 3.41 \\  % 5.39
  \noalign{\vspace{2pt}}
  \hline
  \end{tabular}
\end{table}

\vfill\eject

\section*{Movies}

\noindent Movie S1: Mpeg movie showing the simulation of the mass
ejection during the merger phase. Left panel: particle plot showing
particles that are ejected in red and particles that remain bound in
blue. The scale is in units of 1500$\,R_\odot$.  Right panel:
Histogram of the mass (left axis) ejected as a function of $\sin
\theta$, where $\theta$ is the angle with respect to the equatorial
plane. The thin blue curves give the median velocity (central curve)
and the range of velocities (upper and lower curves) that includes
50\,\% of the ejecta at a given $\sin \theta$. The velocities are
given on the right axis.

\bigskip

\noindent Movie S2: Mpeg movie showing the formation of the
triple-ring nebula as a result of the interaction of the
blue-supergiant wind (grey particles) with the matter ejected in the
merger phase (green/blue particles). The colour of the ejecta
particles indicates the logarithm of the mass density in units of ${\rm
g\,cm}^{-3}$ (see scale bar). The spatial scale is in units of pc.

\clearpage
\bibliographystyle{science}

\begin{thebibliography}{}
\bibitem{som} Materials and methods are available as supporting
material on Science Online.
\end{thebibliography}

\end{document}